\documentclass{pasa}%
\title[Cosmological Inference from Host-Selected Type Ia Supernova Samples]{Cosmological Inference from Host-Selected Type Ia Supernova Samples}
\author[Uddin et al.]{Syed A Uddin$^{1,2,4}$\thanks{Correspondent: saushuvo@gmail.com}, Jeremy Mould$^{1,4}$, Chris Lidman$^{3,4}$, Vanina Ruhlmann-Kleider$^5$, \and Delphine Hardin$^6$  \\
\affil{$^1$Centre for Astrophysics and Supercomputing, Swinburne University of Technology, Melbourne, VIC 3122, Australia}%
\affil{$^2$Purple Mountain Observatory, Chinese Academy of Sciences, Nanjing,, Jiangshu 210008, China}
\affil{$^3$Australian Astronomical Observatory, North Ryde, NSW 2113, Australia}%
\affil{$^4$Australian Research Council Centre of Excellence for All-Sky Astrophysics}%
\affil{$^5$CEA, Centre de Saclay, Irfu/SPP, 91191 Gif-sur-Yvette, Paris, France}
\affil{$^6$Universite Pierre et Marie Curie, Laboratoire de Physique Nucleaire et des Hautes Energies, 4 place Jussieu, Paris, France}}
\jid{PASA}
\doi{10.1017/pas.\the\year.xxx}
\jyear{\the\year}

\usepackage{natbib}
\bibpunct{(}{)}{;}{a}{}{,}
\usepackage{aas_macros}
\usepackage{hyperref} 
\hypersetup{colorlinks,citecolor=blue,linkcolor=blue,urlcolor=blue}

\usepackage{adjustbox}

\begin{document}%
\begin{abstract}

We compare two Type Ia supernova (SN Ia) samples that are drawn from a spectroscopically confirmed SN Ia sample:  a host-selected sample in which SNe Ia are restricted to those that have a spectroscopic redshift from the host; and a broader, more traditional sample in which the redshift could come from either the SN or the host. The host-selected sample is representative of SN samples that will use the redshift of the host to infer the SN redshift, long after the SN has faded from view. We find that SNe Ia that are selected on the availability of a redshift from the host differ from SNe Ia that are from the broader sample. The former tend to be redder, have narrower light curves, live in more massive hosts, and tend to be at lower redshifts. We find that constraints on the equation of state of dark energy, $w$, and the matter density, $\Omega_M$, remain consistent between these two types of samples. Our results are important for ongoing and future supernova surveys, which unlike previous supernova surveys, will have limited real-time follow-up to spectroscopically classify the SNe they discover. Most of the redshifts in these surveys will come from the hosts.

\end{abstract}
\maketitle%
\section{INTRODUCTION }
\label{sec:intro}

Late time cosmic acceleration has been discovered through distance measurements to SNe Ia over a range of redshifts (\citealt{riess98}, \citealt{perlmutter99}). One explanation for this striking discovery is that the universe is dominated by an unknown energy component, termed dark energy, that is driving the observed acceleration. Although the $\Lambda$CDM model of cosmology --  in which $w$, the dark energy equation of state parameter, is exactly -1 -- is consistent with observations, the nature of dark energy remains elusive. Another explanation is that the gravity on very large scales behaves differently to that predicted in General Relativity (see the review by \citealt{joyce16}). Precision cosmological measurements may shed light on the physics that drives the acceleration.

Deriving cosmological parameters using SNe Ia relies on the measurement of two observables: distance and redshift. Distance measurements with a precision of $\sim7\%$ come from the standardisation of SN Ia luminosities (\citealt{phillips93}, \citealt{tripp98}). Recently, $\sim 4\%$ precision in distance measurement has been obtained by using a subset of SNe Ia in high-ultraviolet surface brightness hosts (\citealt{kelly15}). Redshifts can be obtained from either the host or the SN itself.

In past SN surveys, one confirms the SN type by a spectrum of the SN together with its host while the SN is within a couple of weeks of maximum light. In surveys like the Supernova Legacy Survey (SNLS; \citealt{betoule14}; hereafter B14), the amount of time required to do the spectroscopic confirmation was significantly larger than the amount of time used to obtain SN light curves. On the other hand, obtaining the redshift from a spectrum of the host galaxy with or without the SN is simpler and can be done at any time. 

Another advantage of getting the redshift from the host is that the error in the redshift is smaller. The typical error from SN Ia spectra is of the order of $10^{-2}$, whereas the typical error from galaxy spectra is of the order of $10^{-4}$. While redshifts for host galaxies can be determined with greater precision, there are disadvantages. In addition to not having a spectrum to confirm the SN type spectroscopically, the host may be incorrectly identified or too faint to observe \citep{gupta16}.

The next generation of SN surveys will produce many more SNe Ia than past surveys. For example, the ongoing Dark Energy Survey (DES; \citealt{bernstein12}) will discover $\sim$10 times the number of SNe Ia that were discovered in the SNLS. Only a small fraction of the SNe Ia from the new surveys will have spectra of the SNe Ia. We are therefore entering into an era where most SNe Ia will be classified from photometry and most redshifts will be obtained from host galaxies. The efficiency of such a strategy was demonstrated in \citet{lidman13}, where host galaxy redshifts of photometrically classified SNe Ia from the SNLS were obtained. An example of a cosmological analysis using photometrically classified SNe Ia from the SDSS-II supernova survey was presented in \citet{campbell13}. The cosmological inferences that they obtained with their sample are consistent with those obtained from the spectroscopically confirmed SN Ia sample of B14. 


The Dark Energy Survey is also employing a  similar scheme of getting SN Ia redshifts from host galaxies (\citealt{yuan15}).  The SN Ia sample they will produce will extend slightly beyond $z=1$. These SNe Ia will be biased towards galaxies that are bright and actively forming stars as it is easier to get redshifts for these galaxies. A question then arises as to whether or not the cosmological parameters derived using a sample of SNe Ia, where redshifts are obtained from host galaxies, are different from those using a sample of SNe Ia that contains redshifts from both SNe Ia and hosts.

In this paper, we examine this question using the SN Ia sample presented in B14, where information on the source of the redshift is available and SNe Ia are within the redshift range of $0.01<z<1.3$. From this sample we create two types of samples: a host-selected sample, where redshifts come from the host galaxies, and a traditional sample, where the redshift could come from either the SN Ia or its host. 

We organise the paper as follows: In Section \ref{sec:snsample}, we describe the SN Ia sample used in this paper. In Section \ref{sec:compare}, we compare the properties of the SNe Ia in the two samples. In Section \ref{sec:inference}, we derive constraints on cosmological parameters from the two samples and examine the differences. We discuss effect of Malmquist bias in Section \ref{sec:discussion}. We conclude in Section \ref{sec:conclusion}. 


\section{SNe Ia Sample}\label{sec:snsample}
The sample in this work comes from the Joint Light-curve Analysis (JLA; B14). JLA has compiled 740 spectroscopically confirmed SNe Ia from various supernova surveys including SDSS-II (\citealt{frieman08}, \citealt{holtzman08}, \citealt{zheng08}, \citealt{sako11}, and \citealt{sako14}), SNLS (\citealt{howell05}, \citealt{bronder08}, \citealt{ellis08}, \citealt{balland09}, and \citealt{walker11}), CfA (\citealt{riess99}, \citealt{jha07}, \citealt{hicken09}, and \citealt{hicken12}), CSP (\citealt{hamuy06}, \citealt{contreras10}, and \citealt{stritzinger11}), and HST (\citealt{riess07}). Light-curves of all the SNe Ia in this sample are fitted using the SALT\footnote{http://supernovae.in2p3.fr/salt/doku.php} (\citealt{guy10}; version 2.4.0) light-curve fitter, as it is the most widely used and tested light-curve fitter in the literature. Of these 740 SNe Ia, there are 537 SNe Ia for which redshifts were derived from the hosts. This means that there were 203 SNe Ia in which the SN Ia redshift could only be derived from the SN Ia itself. Table~\ref{sample} shows the JLA sample with redshift source information. Throughout the rest of this paper we will refer to the sample of SNe Ia whose redshifts come from either the host or the SN Ia as the traditional sample, and we will refer to the sample  whose SNe Ia are restricted to those that have a spectroscopic redshift from the host as the host-selected sample.

\section{Comparing Traditional and Host-selected Samples}\label{sec:compare}

In this section we want to compare the properties of SNe Ia between traditional and host-selected samples. The JLA sample of 740 SNe Ia described above can act as a traditional sample and the sub-sample of 537 SNe Ia can act as a host-selected sample. But there is $73\%$ overlap between these two samples. Therefore, any statistical comparison between these two is not instructive. We need to develop a way to create traditional and host-selected samples so that there is no overlap between them.

A way to create two independent samples is to take the JLA sample and split it into two equal samples of 370 SNe Ia each in such a way that one sample is a traditional sample and the other is a host-selected sample. First, we divide the JLA sample into low-redshift and high-redshift groups. Low-redshift SNe Ia are from galaxy targeted surveys. Their redshifts are from host galaxies and hence host-selected. We take all low-redshift SNe Ia and randomly select (without replacement) half of them and put in a box labelled A. We put the other half of the low-redshift SNe Ia in a box labelled B. Next, we take the high-redshift SNe Ia and select those SNe Ia that lack a redshift from the host (i.e. the redshift comes from the SN itself). We put them in box A. Now what remains are the high-redshift SNe Ia that have redshifts from the host. We then randomly distribute them into box A and box B in such a way that the total number of SNe Ia in each box becomes 370. Now box A becomes the traditional sample and box B becomes the host-selected sample. The disadvantage of this method is that the traditional sample contains more SNe Ia ($55\%$ of the total) that have redshifts from the SNe Ia themselves than would normally be the case. 

\begin{figure*}[htbp]
\begin{center}
\includegraphics[width=1.0\textwidth]{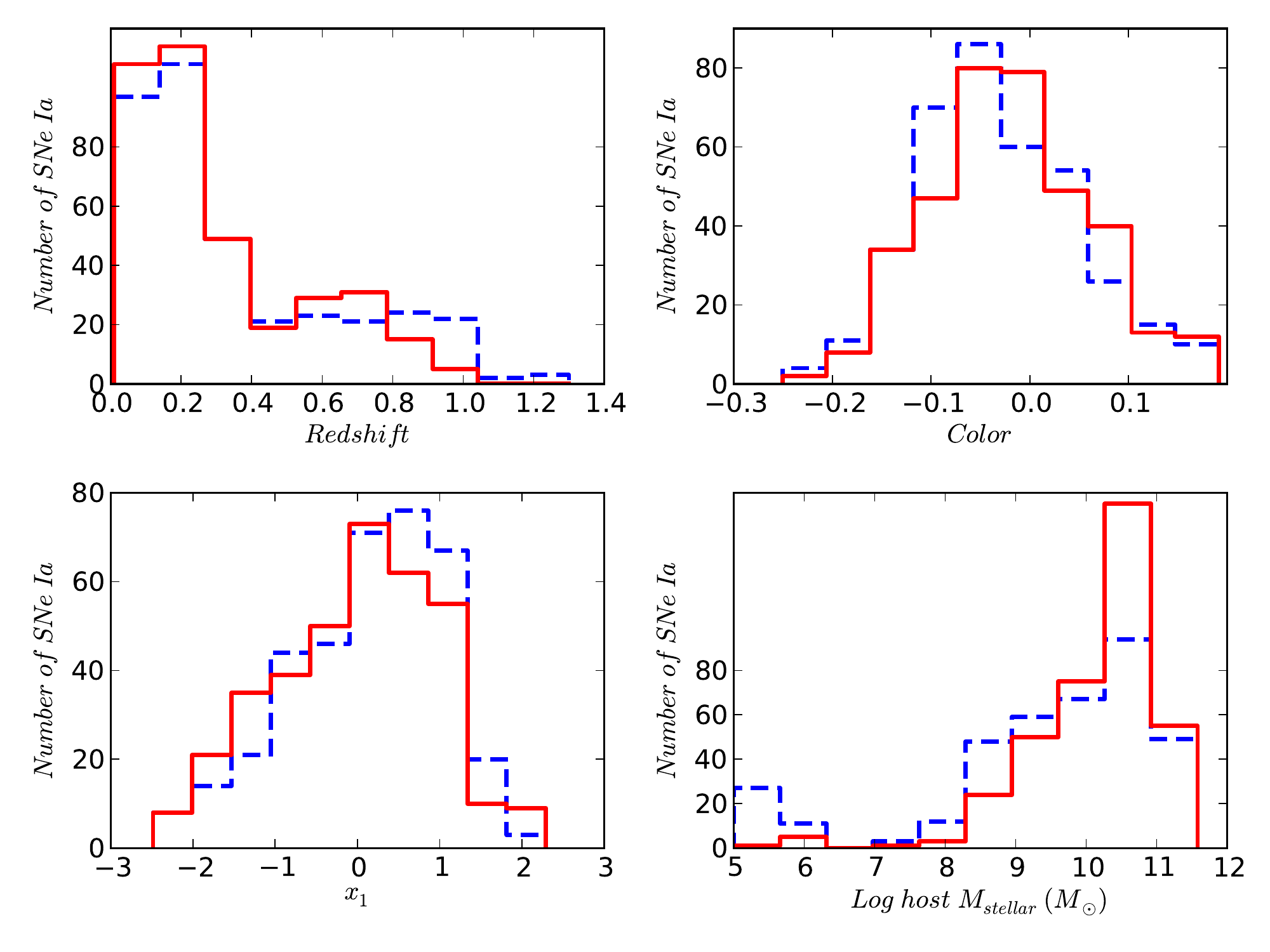}
\caption{Distribution of SN Ia redshift (top left), colour (top right), stretch (bottom left), and mean host stellar mass (bottom right) for both the host-selected (solid red) and traditional (dashed blue) samples. These distributions are shown for one random realisation.}
\label{onedist}
\end{center}
\end{figure*}

In Fig.~\ref{onedist} we show the distributions of redshifts, colour, stretch, and host mass for these two types of samples for one random realisation. We see that SNe Ia in the traditional sample tend to occur at higher redshift, are bluer, and have narrower light-curve widths. Galaxies in the host-selected sample also tend to be more massive.

\begin{figure*}[htbp]
\begin{center}
\includegraphics[width=1.0\textwidth]{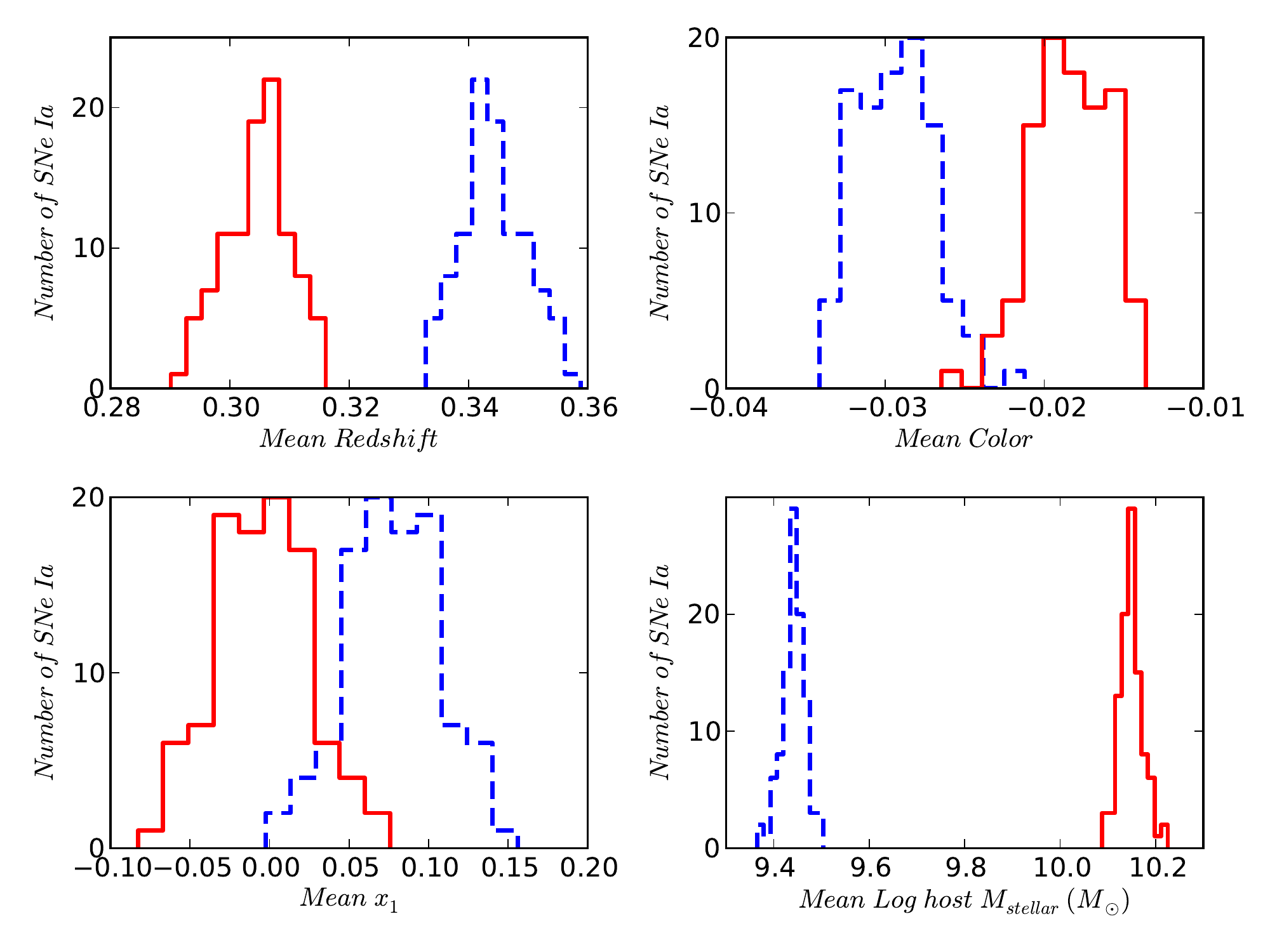}
\caption{Distribution of SN Ia mean redshift (top left), mean colour (top right), mean stretch (bottom left), and mean host stellar mass (bottom right) for host-selected (solid red) and traditional (dashed blue) samples. Note how the two distributions differ for each property. See text for details. The mirror symmetry in the figures is a consequence of the way SNe Ia are split between the two samples. The SNe Ia are selected without replacement. This and the even number of SNe Ia in the two subsamples leads to the mirror symmetry.}
\label{zx1c_z}
\end{center}
\end{figure*}

We now compare traditional and host-selected samples by calculating mean properties of SNe Ia and their hosts in each sample. We repeat the process of creating traditional and host-selected samples 100 times, calculating the mean properties of SNe Ia and their hosts of each sample. Fig.~\ref{zx1c_z} shows the distributions of SN Ia mean redshift, mean colour, mean stretch, and mean host stellar mass for both the traditional and the host-selected samples.
 


It is interesting to see from Fig.~\ref{zx1c_z} that the distributions of the mean properties are different between two types of samples. SNe Ia in the traditional sample are, on average, bluer and broader. They also have a higher mean redshift. This is not unexpected, as it is harder to get redshifts for galaxies that are further away. However, there may be evolutionary effects that could either drive or mitigate the difference. For example, the number of SNe Ia in star forming galaxies relative to the number of SNe Ia in passive galaxies is expected to increase with redshift (\citealt{mannucci06}, \citealt{childress13}). This would tend to mitigate the difference between the two samples as it is generally easier to measure the redshift of a galaxy that is actively star-forming. We also find that the mean host stellar mass in host-selected samples are higher than the traditional samples. This is because it is easy to obtain redshifts of galaxies that are bright and hence more massive.

\section{Cosmological Inferences}\label{sec:inference}

We have shown that SNe Ia in the host-selected samples are different to traditional samples in terms of redshift, colour, stretch, and host mass. We now examine if the cosmological parameters derived from these samples differ. For this we perform fits to both the $\Lambda$CDM and $w$CDM models to the 100 pairs of randomly created traditional host-selected samples.

We compute the observed distance modulus using the fitted SN Ia rest frame $B$-band peak apparent magnitude ($m_B$), corrected for stretch ($x_1$) and colour ($c$). This correction is needed for standardising SNe Ia as distance indicators (\citealt{phillips93}, \citealt{tripp98}, \citealt{astier06}, \citealt{goobar11}).  The observed distance modulus is given by: 



\small
\begin{equation}
\mu_o =
\begin{cases} 
m_B + \alpha x_1 -\beta c -M_B & M_{host}< 10^{10}M_{\odot} \\
m_B + \alpha x_1 -\beta c -M_B +\Delta M_B& M_{host} \geq 10^{10}M_{\odot} 
\end{cases} 
\end{equation}

\normalsize
\noindent
The parameters that describe the relationships between SN Ia luminosity  ($M_B$), stretch, and colour ($\alpha$ and $\beta$) are derived during the cosmological fit. Here, we have used a host mass correction $(\Delta M_B)$ as suggested by \cite{sullivan10}. It is known that the average SN Ia luminosity after correcting for light curve width and colour correlates with the host mass (e.g., \citealt{sullivan10}, \citealt{childress13}, \citealt{uddinphd}) in such a way that massive galaxies host more luminous SNe Ia.  Unaccounted for, effects like this might bias the accurate measurement of the properties of dark energy.

For each SN Ia, we compute the model distance modulus using :
\begin{equation}
\mu_m = 5\log_{10}d_L (\Omega_M, w)+25 
\end{equation}
 
\noindent
Here the cosmological dependence resides in luminosity distance $d_L$. We define the Hubble residual as $\Delta \mu \equiv \mu_o-\mu_m$. Following B14, we fit the cosmological parameters ($\Omega_m$, $w$) along with nuisance variables ($\alpha$, $\beta$, $M_B$) by minimising

\begin{equation}
\bf{\chi^2 = \Delta \mu^{\dagger}C^{-1} \Delta \mu}
\label{chi}
\end{equation}

\noindent
where $\bf{C}$ is the covariance matrix that includes all statistical and systematic uncertainties, including terms due to peculiar velocities, gravitational lensing, and intrinsic scatter. The matrix \textbf{C} is defined as:

\small
\begin{equation}
\bf C = AC_{\eta}A^{\dagger} + diag \bigg(\frac{5\sigma_z}{z\log_{10}}\bigg)^2+diag(\sigma^2_{\rm{lens}})+diag(\sigma^2_{\rm{int}})
\label{covar}
\end{equation}

\normalsize

\noindent
where $\bf{C_{\eta}}$ is the covariance matrix defined in B14.  $\sigma_z$ in the second term of Eqn.~\ref{covar} is due to peculiar velocities and is set to $c\sigma_z=\mathrm{150\ kms^{-1}}$ \citep{conley11}. The third term is due to gravitational lensing, which varies with redshift according to as $\sigma_{\rm{lens}}=0.055\times z$ \citep{jonsson10}.

The last term in Eqn.~\ref{covar} incorporates the intrinsic dispersion in corrected SN Ia luminosities, $\sigma_{int}$. This term is needed as SNe Ia are not perfect standard candles. We do not know what causes the extra dispersion in their luminosities \citep{conley11}. B14 calculates redshift dependent $\sigma_{int}$ for each survey and we use those values here.

Seven different types of systematic uncertainties are considered in B14: uncertainties in the photometric calibration, light-curve model uncertainty, uncertainties in the Malmquist bias correction, mass step error, uncertainties in the extinction correction from Galactic dust, uncertainties in the mean amplitude of peculiar velocities, and contamination from non-SNe Ia. 

In our analysis, we assume that the Malmquist bias correction and the associated uncertainty are the same for both the traditional and the host-selected samples. In Section \ref{mbias} we show that this assumption does not make any significant difference in our results. We use the corrections as used in B14, originally calculated in \citep{mosher14}. The corrections were calculated for each survey and they are redshift dependent. The mean correction in the JLA sample  is -0.004 mag with minimum and maximum corrections as -0.036 mag and 0.002 mag respectively.

We use Markov Chain Monte Carlo (MCMC) to explore the likelihood defined in Eqn.~\ref{chi} and derive best-fit  parameters. We use the MCMC sampler called EMCEE\footnote{http://dan.iel.fm/emcee/} \citep{foreman13}. EMCEE is a Python implementation of an affine-invariant ensemble sampler. We fit both the $\Lambda$CDM and the $w$CDM models to the traditional and host-selected samples. 

\begin{figure}[htbp]
\begin{center}
\includegraphics[width=.5\textwidth]{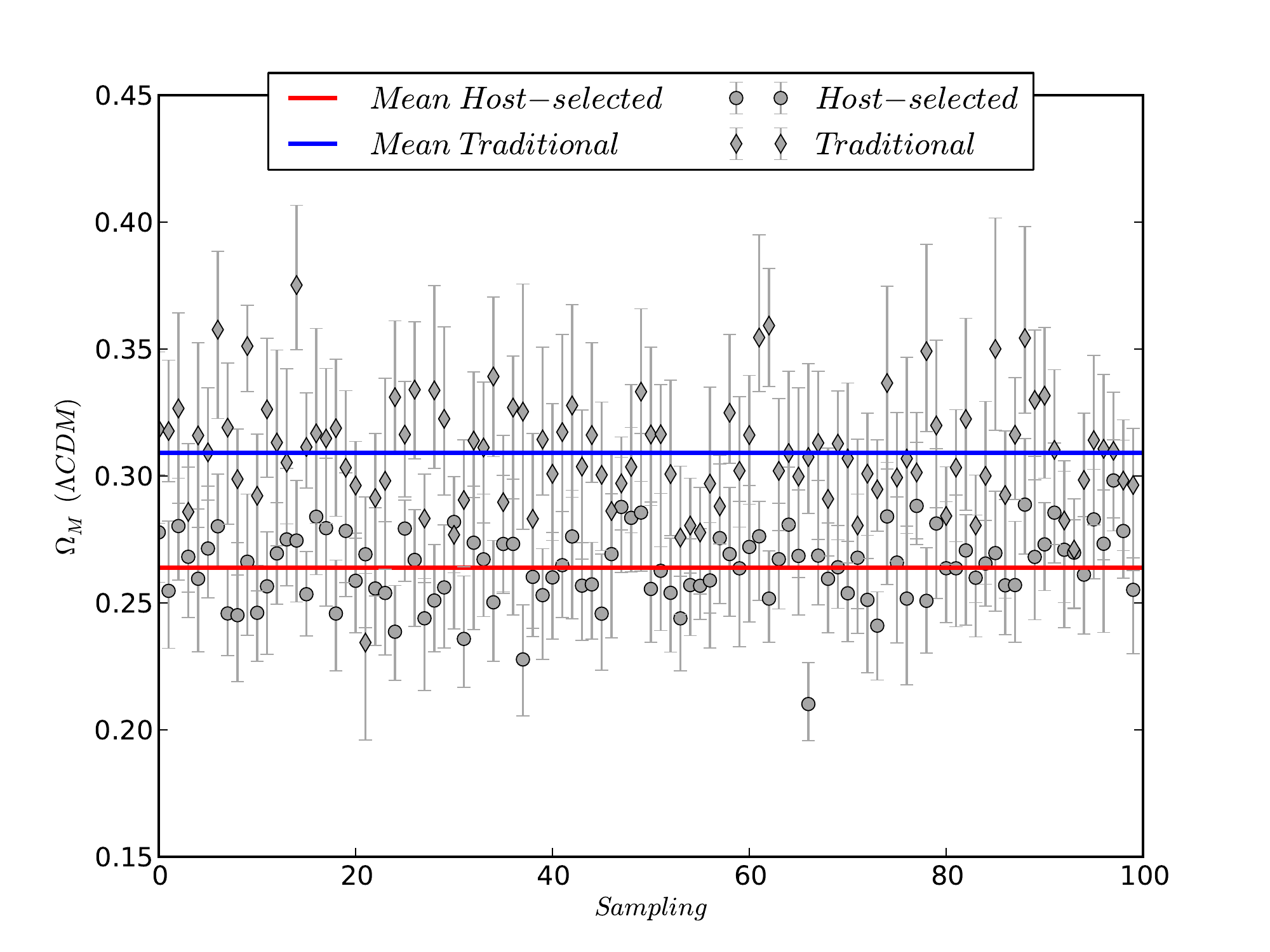}
\caption{Values of $\Omega_M$ in the $\Lambda$CDM model obtained from 100 random realisations of host-selected and traditional samples. Weighted means are also shown. Although there is a shift in $\Omega_M$ of 0.046 between these two samples, the shift is not significant (see Table \ref{result} and the discussion in Section \ref{sec:random}).}
\label{sample_lcdm}
\end{center}
\end{figure}

\begin{figure}[htbp]
\begin{center}
\includegraphics[width=.5\textwidth]{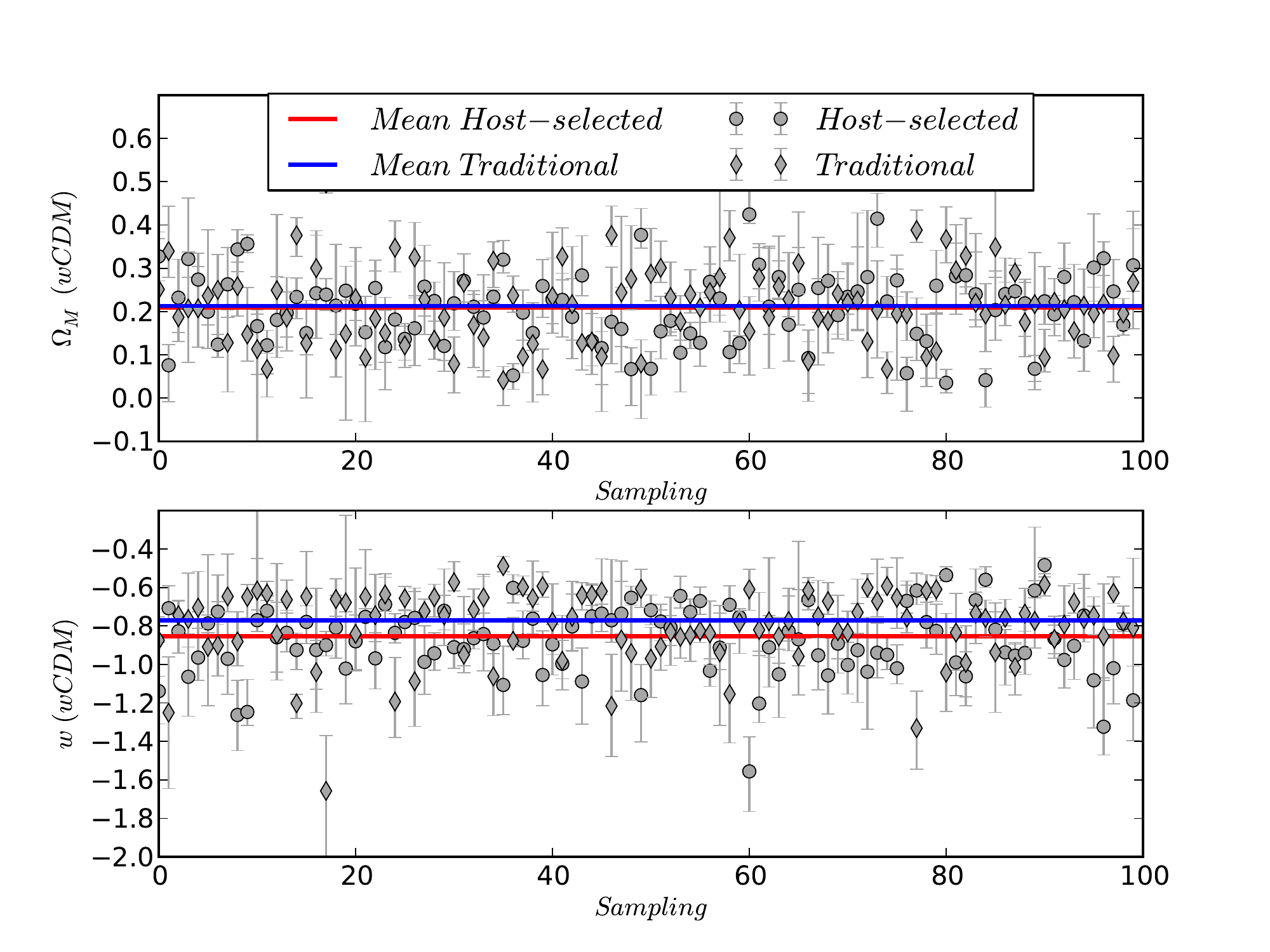}
\caption{Values $\Omega_M$ (top) and $w$ (bottom) in the $w$CDM model obtained from 100 random realisations of host-selected and traditional samples. Weighted means are also shown. There are no significant offsets in $w$ and $\Omega_M$ between these two samples. }
\label{sample_wcdm}
\end{center}
\end{figure}

In comparing the results, we follow the procedure described in the previous section. We create 100 pairs of samples and plot best-fit cosmological parameters for each pair. These are shown in Fig.~\ref{sample_lcdm} for $\Lambda$CDM and Fig.~\ref{sample_wcdm} for $w$CDM respectively along with their weighted means when full covariance matrix is used. Table~\ref{result} summarises mean values and median errors  of the cosmological parameters from these samples.

\begin{table*}
\caption{Weighted mean and $1\sigma$ error of the 100 sampling.}
\begin{center}
\begin{tabular}{llccccc}
\hline
Cosmology& Sample &Parameter& Mean  &Median error  &Mean  &Median error \\
&&&(stat+sys) &(stat+sys) &(stat) &(stat)\\
\hline
$\Lambda$CDM & Host-Selected& $\Omega_M$ & 0.263 & 0.034 & 0.261 & 0.029\\
$\Lambda$CDM& Traditional & $\Omega_M$& 0.309 & 0.045 & 0.264 & 0.034\\
$w$CDM &Host-Selected & $\Omega_M$& 0.209 &0.097 & 0.226 & 0.086\\
$w$CDM &Traditional & $\Omega_M$& 0.213 &0.115 & 0.206 & 0.096\\
$w$CDM &Host-Selected & $w$& -0.852 &0.186 & -0.877 & 0.181\\
$w$CDM &Traditional & $w$& -0.769 &0.194 & -0.749 & 0.171\\
\hline
\end{tabular}
\end{center}
\label{result}
\end{table*}%


For both cosmological models, we do not find any significant tension in the cosmological parameters, $\Omega_M$ and $w$, between the host-selected and traditional samples. The value of $\Omega_M$ in the $\Lambda$CDM model differs by 0.046 between the samples, but this shift is insignificant (0.8$\sigma$). These results indicate that cosmological constraints from future SN surveys may remain unbiased in comparison with the current surveys. We check the robustness of this finding in Section \ref{sec:random}.

We note that the systematic uncertainties of the two randomly drawn samples will be correlated. We investigate the impact this has on our results by  redoing the analysis without including systematic uncertainties. The results are listed in Table~\ref{result}. The impact is minor and our conclusions are unchanged.

\section{Discussion}\label{sec:discussion}

\subsection{Effect of Malmquist Bias}\label{mbias}

We have mentioned in Section \ref{sec:inference} that we have kept the Malmquist bias unchanged for both types of samples. This is unlikely to be true in detail. To see if a change in the magnitude of the Malmquist bias could change our results, we alter the Malmquist bias correction for the host-selected sample and rerun the analysis as we do in Section \ref{sec:inference}. 

When we double the Malmquist bias correction for the host-selected sample, $\Omega_M$ changes by less than $1\%$, which is smaller than our error. Excluding it altogether results in an even smaller change. Our results are therefore not sensitive to the level of Malmquist bias that can be expected for SN samples. Hence, our choice in setting the magnitude of the bias to be the same for both subsamples is reasonable.

\subsection{The significance of the offset in $\Omega_M$}\label{sec:random}
We have seen in Section \ref{sec:inference} that there was an offset in $\Omega_M$ of 0.046 between the two samples in $\Lambda$CDM. Using the median error of the two samples, we determined that the offset was not significant. Here we re-examine the significance of the offset using random subsamples.

\begin{figure*}[!b]
\centering
\begin{tabular}{ccc}
\includegraphics[width=0.5\textwidth]{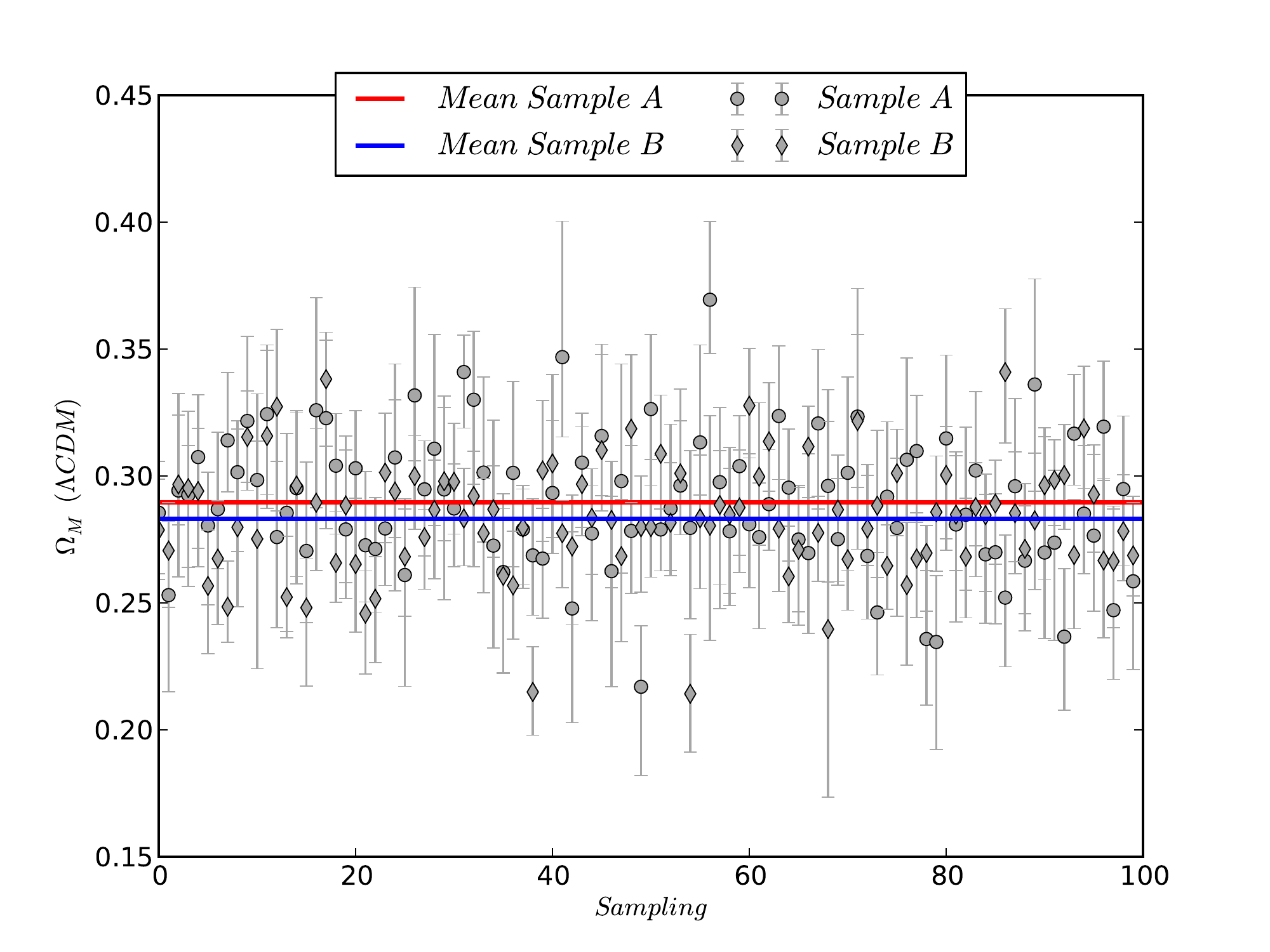} &
\includegraphics[width=0.5\textwidth]{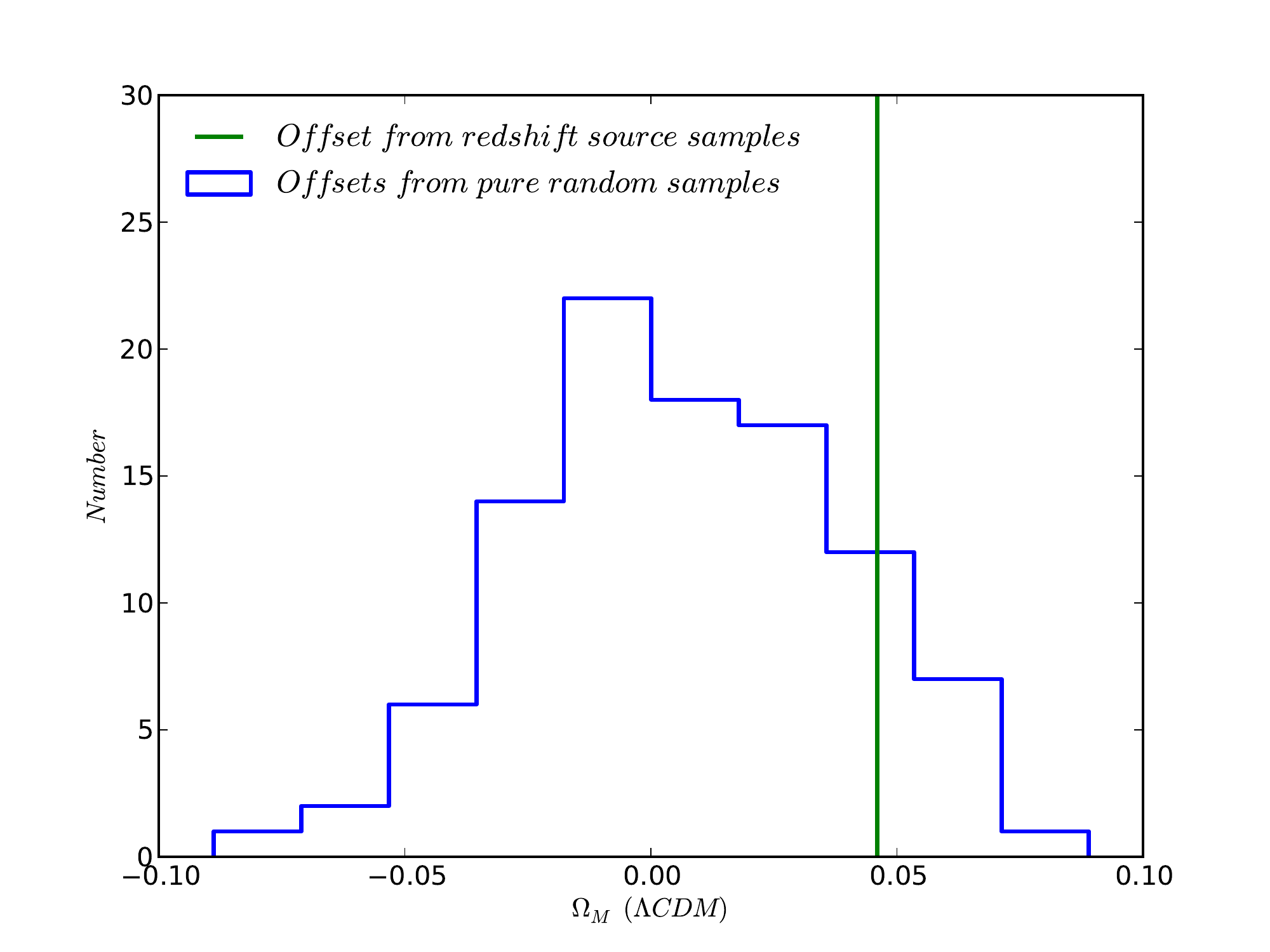}
\end{tabular}
\caption{Left: Values of $\Omega_M$ in the $\Lambda$CDM model obtained from 100 realisations. Weighted means are also shown. We find no significant offset between these two samples. Right: Distribution of the offsets in $\Omega_M$ for each pair of A and B. Also shown is the offset from the wighted mean of Fig.~\ref{sample_lcdm} - which is 0.046. We see that the offset in $\Omega_M$ between samples A and B exceeds the mean offset between the host-selected and traditional samples $12\%$ of the time. }
\label{pure}
\end{figure*}

To create these subsamples we take the JLA sample and split it into two equal halves with the constraint that the ratio of low and high redshift SNe Ia in the two subsamples is the same. Let us call these two random samples A and B. We then calculate cosmological parameter in the $\Lambda$CDM model. We repeat this 100 times and calculate weighted means in each group of A and B. We show these results in Fig.~\ref{pure}.

We find no significant offset in $\Omega_M$ between the samples (see Fig.~\ref{pure}, left). We then ask how many times the offset in $\Omega_M$ exceeds the mean offset between host-selected and traditional samples. To answer this, we plot the distribution of the offset in $\Omega_M$ from 100 realisations (see Fig.~\ref{pure}, right). We also  mark the mean offset of 0.046 that we have found in Section \ref{sec:inference} (also in Fig.~\ref{sample_lcdm}). We find that the offset in $\Omega_M$ between samples A and B exceeds the mean offset between the host-selected and traditional samples $12\%$ of the time.

\section{Conclusion}\label{sec:conclusion}

In this paper, we have compared SN Ia properties and cosmological parameters from a traditional SN Ia sample, where redshifts come from either the SNe Ia or the galaxies that host them, and a host-selected SN Ia sample, where redshifts come entirely from the host galaxies. The host selected SN Ia sample is similar to the samples that surveys such as DES are currently obtaining, where most of the redshifts will come from SN host galaxies.

We have found that the mean redshift, colour, stretch, and host galaxy stellar mass of SNe Ia can be different between these two types of samples. Using SNe Ia alone, we find that the two subsamples give consistent cosmological results. In future work, we will examine if this result continues to be true when cross-cutting cosmological constraints from other probes are added.

Future SN Ia samples, such as the sample of SNe Ia from DES, will be several times larger. While extending the analysis presented in this paper to these larger samples will be limited by the number of SNe Ia that are spectroscopically confirmed in these samples (expected to be 20$\%$ of the total), our analysis will be an important one to make in these future surveys.

\begin{acknowledgements}
We thank Chris Blake for helpful discussion regarding statistical analysis.  We also thank Tamara Davis and Christophe Balland for their valuable comments. Part of this research was conducted by the Australian Research Council Centre of Excellence for All-sky Astrophysics (CAASTRO), through project number CE110001020. This research made use of Astropy, a community-developed core Python package for Astronomy \citep{astropy13}.
\end{acknowledgements}

\begin{appendix}

\section{Table of SN Ia Redshift Sources}

\begin{table}[htp]
\caption{Redshift source for SNe Ia in the JLA sample. The full list is available online.}
\begin{center}
\begin{tabular}{ll}
\hline
SN Ia & Redshift Source\\
\hline
03D1fc	&	Host\\
03D1fq	&	SN\\
03D3ba	&	Host\\
03D4cj	&	SN\\
03D4cx	&	SN\\
06D4dh	&	Host\\
Aphrodite	&	SN\\
Eagle	&	SN\\
Elvis	&	SN\\
Patuxent	&	SN\\
SDSS12781	&	Host\\
SDSS12843	&	Host\\
SDSS12853	&	SN\\
SDSS12855	&	SN\\
SDSS9457	&	Host\\
SDSS9467	&	Host\\
sn2006ob	&	Host\\
sn2006on	&	Host\\
sn2006qo	&	Host\\
\hline
\end{tabular}
\end{center}
\label{sample}
\end{table}%

\end{appendix}

\bibliographystyle{pasa-mnras}
\bibliography{bibSyedUddin}

\begin{thebibliography}{}
\makeatletter
\relax
\def\mn@urlcharsother{\let\do\@makeother \do\$\do\&\do\#\do\^\do\_\do\%\do\~}
\definecolor{darkblue}{rgb}{0,0,0.597656}
\def\mndoi{\begingroup\mn@urlcharsother \@ifnextchar [ {\mndoi@} {\mndoi@[]}}
\def\mndoi@[#1]#2{\def\@tempa{#1}\ifx\@tempa\@empty \href
  {http://dx.doi.org/#2} {\textcolor{darkblue}{doi:#2}}\else \href
  {http://dx.doi.org/#2} {\textcolor{darkblue}{#1}}\fi \endgroup}
\def\mn@eprint#1#2{\mn@eprint@#1:#2::\@nil}
\def\mn@eprint@arXiv#1{\href {http://arxiv.org/abs/#1} {{\tt arXiv:#1}}}
\def\mn@eprint@dblp#1{\href {http://dblp.uni-trier.de/rec/bibtex/#1.xml}
  {dblp:#1}}
\def\mn@eprint@#1:#2:#3:#4\@nil{\def\@tempa {#1}\def\@tempb {#2}\def\@tempc
  {#3}\ifx \@tempc \@empty \let \@tempc \@tempb \let \@tempb \@tempa \fi \ifx
  \@tempb \@empty \def\@tempb {arXiv}\fi \@ifundefined
  {mn@eprint@\@tempb}{\@tempb:\@tempc}{\expandafter \expandafter \csname
  mn@eprint@\@tempb\endcsname \expandafter{\@tempc}}}

\bibitem[\protect\citeauthoryear{{Astropy Collaboration} et~al.,}{{Astropy
  Collaboration} et~al.}{2013}]{astropy13}
{Astropy Collaboration} et~al., 2013, \mndoi [\aap]
  {10.1051/0004-6361/201322068}, \href
  {http://adsabs.harvard.edu/abs/2013A%26A...558A..33A} {558, A33}

\bibitem[\protect\citeauthoryear{{Bernstein} et~al.,}{{Bernstein}
  et~al.}{2012}]{bernstein12}
{Bernstein} J.~P.,  et~al., 2012, \mndoi [\apj] {10.1088/0004-637X/753/2/152},
  \href {http://adsabs.harvard.edu/abs/2012ApJ...753..152B} {753, 152}

\bibitem[\protect\citeauthoryear{{Betoule} et~al.,}{{Betoule}
  et~al.}{2014}]{betoule14}
{Betoule} M.,  et~al., 2014, \mndoi [\aap] {10.1051/0004-6361/201423413}, \href
  {http://adsabs.harvard.edu/abs/2014A%26A...568A..22B} {568, A22}

\bibitem[\protect\citeauthoryear{{Campbell} et~al.,}{{Campbell}
  et~al.}{2013}]{campbell13}
{Campbell} H.,  et~al., 2013, \mndoi [\apj] {10.1088/0004-637X/763/2/88}, \href
  {http://adsabs.harvard.edu/abs/2013ApJ...763...88C} {763, 88}

\bibitem[\protect\citeauthoryear{{Childress} et~al.,}{{Childress}
  et~al.}{2013}]{childress13}
{Childress} M.,  et~al., 2013, \mndoi [\apj] {10.1088/0004-637X/770/2/108},
  \href {http://adsabs.harvard.edu/abs/2013ApJ...770..108C} {770, 108}

\bibitem[\protect\citeauthoryear{{Conley} et~al.,}{{Conley}
  et~al.}{2011}]{conley11}
{Conley} A.,  et~al., 2011, \mndoi [\apjs] {10.1088/0067-0049/192/1/1}, \href
  {http://adsabs.harvard.edu/abs/2011ApJS..192....1C} {192, 1}

\bibitem[\protect\citeauthoryear{{Contreras} et~al.,}{{Contreras}
  et~al.}{2010}]{contreras10}
{Contreras} C.,  et~al., 2010, \mndoi [\aj] {10.1088/0004-6256/139/2/519},
  \href {http://adsabs.harvard.edu/abs/2010AJ....139..519C} {139, 519}

\bibitem[\protect\citeauthoryear{{Foreman-Mackey}, {Hogg}, {Lang}  \&
  {Goodman}}{{Foreman-Mackey} et~al.}{2013}]{foreman13}
{Foreman-Mackey} D.,  {Hogg} D.~W.,  {Lang} D.,   {Goodman} J.,  2013, \mndoi
  [\pasp] {10.1086/670067}, \href
  {http://adsabs.harvard.edu/abs/2013PASP..125..306F} {125, 306}

\bibitem[\protect\citeauthoryear{{Goobar} \& {Leibundgut}}{{Goobar} \&
  {Leibundgut}}{2011}]{goobar11}
{Goobar} A.,  {Leibundgut} B.,  2011, \mndoi [Annual Review of Nuclear and
  Particle Science] {10.1146/annurev-nucl-102010-130434}, \href
  {http://adsabs.harvard.edu/abs/2011ARNPS..61..251G} {61, 251}

\bibitem[\protect\citeauthoryear{{Gupta} et~al.,}{{Gupta}
  et~al.}{2016}]{gupta16}
{Gupta} R.~R.,  et~al., 2016, preprint, \href
  {http://adsabs.harvard.edu/abs/2016arXiv160406138G} {} (\mn@eprint {arXiv}
  {1604.06138})

\bibitem[\protect\citeauthoryear{{Hicken} et~al.,}{{Hicken}
  et~al.}{2009}]{hicken09}
{Hicken} M.,  et~al., 2009, \mndoi [\apj] {10.1088/0004-637X/700/1/331}, \href
  {http://adsabs.harvard.edu/abs/2009ApJ...700..331H} {700, 331}

\bibitem[\protect\citeauthoryear{{J{\"o}nsson} et~al.,}{{J{\"o}nsson}
  et~al.}{2010}]{jonsson10}
{J{\"o}nsson} J.,  et~al., 2010, \mndoi [\mnras]
  {10.1111/j.1365-2966.2010.16467.x}, \href
  {http://adsabs.harvard.edu/abs/2010MNRAS.405..535J} {405, 535}

\bibitem[\protect\citeauthoryear{{Joyce}, {Lombriser}  \& {Schmidt}}{{Joyce}
  et~al.}{2016}]{joyce16}
{Joyce} A.,  {Lombriser} L.,   {Schmidt} F.,  2016, preprint, \href
  {http://adsabs.harvard.edu/abs/2016arXiv160106133J} {} (\mn@eprint {arXiv}
  {1601.06133})

\bibitem[\protect\citeauthoryear{{Lidman} et~al.,}{{Lidman}
  et~al.}{2013}]{lidman13}
{Lidman} C.,  et~al., 2013, \mndoi [\pasa] {10.1017/pasa.2012.001}, \href
  {http://adsabs.harvard.edu/abs/2013PASA...30....1L} {30, 1}

\bibitem[\protect\citeauthoryear{{Perlmutter} et~al.,}{{Perlmutter}
  et~al.}{1999}]{perlmutter99}
{Perlmutter} S.,  et~al., 1999, \mndoi [\apj] {10.1086/307221}, \href
  {http://adsabs.harvard.edu/abs/1999ApJ...517..565P} {517, 565}

\bibitem[\protect\citeauthoryear{{Phillips}}{{Phillips}}{1993}]{phillips93}
{Phillips} M.~M.,  1993, \mndoi [\apjl] {10.1086/186970}, \href
  {http://adsabs.harvard.edu/abs/1993ApJ...413L.105P} {413, L105}

\bibitem[\protect\citeauthoryear{{Riess} et~al.,}{{Riess}
  et~al.}{1998}]{riess98}
{Riess} A.~G.,  et~al., 1998, \mndoi [\aj] {10.1086/300499}, \href
  {http://adsabs.harvard.edu/abs/1998AJ....116.1009R} {116, 1009}

\bibitem[\protect\citeauthoryear{{Riess} et~al.,}{{Riess}
  et~al.}{2007}]{riess07}
{Riess} A.~G.,  et~al., 2007, \mndoi [\apj] {10.1086/510378}, \href
  {http://adsabs.harvard.edu/abs/2007ApJ...659...98R} {659, 98}

\bibitem[\protect\citeauthoryear{{Sako} et~al.,}{{Sako} et~al.}{2014}]{sako14}
{Sako} M.,  et~al., 2014, preprint, \href
  {http://adsabs.harvard.edu/abs/2014arXiv1401.3317S} {} (\mn@eprint {arXiv}
  {1401.3317})

\bibitem[\protect\citeauthoryear{{Sullivan} et~al.,}{{Sullivan}
  et~al.}{2010}]{sullivan10}
{Sullivan} M.,  et~al., 2010, \mndoi [\mnras]
  {10.1111/j.1365-2966.2010.16731.x}, \href
  {http://adsabs.harvard.edu/abs/2010MNRAS.406..782S} {406, 782}

\bibitem[\protect\citeauthoryear{{Tripp}}{{Tripp}}{1998}]{tripp98}
{Tripp} R.,  1998, \aap, \href
  {http://adsabs.harvard.edu/abs/1998A%26A...331..815T} {331, 815}

\bibitem[\protect\citeauthoryear{Uddin}{Uddin}{2016}]{uddinphd}
Uddin S.~A.,  2016, PhD thesis, Swinburne University of Technology

\bibitem[\protect\citeauthoryear{{Yuan} et~al.,}{{Yuan} et~al.}{2015}]{yuan15}
{Yuan} F.,  et~al., 2015, \mndoi [\mnras] {10.1093/mnras/stv1507}, \href
  {http://adsabs.harvard.edu/abs/2015MNRAS.452.3047Y} {452, 3047}

\makeatother
\end{thebibliography}


\begin{thebibliography}{}
\makeatletter
\relax
\def\mn@urlcharsother{\let\do\@makeother \do\$\do\&\do\#\do\^\do\_\do\%\do\~}
\definecolor{darkblue}{rgb}{0,0,0.597656}
\def\mndoi{\begingroup\mn@urlcharsother \@ifnextchar [ {\mndoi@} {\mndoi@[]}}
\def\mndoi@[#1]#2{\def\@tempa{#1}\ifx\@tempa\@empty \href
  {http://dx.doi.org/#2} {\textcolor{darkblue}{doi:#2}}\else \href
  {http://dx.doi.org/#2} {\textcolor{darkblue}{#1}}\fi \endgroup}
\def\mn@eprint#1#2{\mn@eprint@#1:#2::\@nil}
\def\mn@eprint@arXiv#1{\href {http://arxiv.org/abs/#1} {{\tt arXiv:#1}}}
\def\mn@eprint@dblp#1{\href {http://dblp.uni-trier.de/rec/bibtex/#1.xml}
  {dblp:#1}}
\def\mn@eprint@#1:#2:#3:#4\@nil{\def\@tempa {#1}\def\@tempb {#2}\def\@tempc
  {#3}\ifx \@tempc \@empty \let \@tempc \@tempb \let \@tempb \@tempa \fi \ifx
  \@tempb \@empty \def\@tempb {arXiv}\fi \@ifundefined
  {mn@eprint@\@tempb}{\@tempb:\@tempc}{\expandafter \expandafter \csname
  mn@eprint@\@tempb\endcsname \expandafter{\@tempc}}}

\bibitem[\protect\citeauthoryear{{Astier} et~al.,}{{Astier}
  et~al.}{2006}]{astier06}
{Astier} P.,  et~al., 2006, \mndoi [\aap] {10.1051/0004-6361:20054185}, \href
  {http://adsabs.harvard.edu/abs/2006A%26A...447...31A} {447, 31}

\bibitem[\protect\citeauthoryear{{Astropy Collaboration} et~al.,}{{Astropy
  Collaboration} et~al.}{2013}]{astropy13}
{Astropy Collaboration} et~al., 2013, \mndoi [\aap]
  {10.1051/0004-6361/201322068}, \href
  {http://adsabs.harvard.edu/abs/2013A%26A...558A..33A} {558, A33}

\bibitem[\protect\citeauthoryear{{Balland} et~al.,}{{Balland}
  et~al.}{2009}]{balland09}
{Balland} C.,  et~al., 2009, \mndoi [\aap] {10.1051/0004-6361/200912246}, \href
  {http://adsabs.harvard.edu/abs/2009A%26A...507...85B} {507, 85}

\bibitem[\protect\citeauthoryear{{Bernstein} et~al.,}{{Bernstein}
  et~al.}{2012}]{bernstein12}
{Bernstein} J.~P.,  et~al., 2012, \mndoi [\apj] {10.1088/0004-637X/753/2/152},
  \href {http://adsabs.harvard.edu/abs/2012ApJ...753..152B} {753, 152}

\bibitem[\protect\citeauthoryear{{Betoule} et~al.,}{{Betoule}
  et~al.}{2014}]{betoule14}
{Betoule} M.,  et~al., 2014, \mndoi [\aap] {10.1051/0004-6361/201423413}, \href
  {http://adsabs.harvard.edu/abs/2014A%26A...568A..22B} {568, A22}

\bibitem[\protect\citeauthoryear{{Bronder} et~al.,}{{Bronder}
  et~al.}{2008}]{bronder08}
{Bronder} T.~J.,  et~al., 2008, \mndoi [\aap] {10.1051/0004-6361:20077655},
  \href {http://adsabs.harvard.edu/abs/2008A%26A...477..717B} {477, 717}

\bibitem[\protect\citeauthoryear{{Campbell} et~al.,}{{Campbell}
  et~al.}{2013}]{campbell13}
{Campbell} H.,  et~al., 2013, \mndoi [\apj] {10.1088/0004-637X/763/2/88}, \href
  {http://adsabs.harvard.edu/abs/2013ApJ...763...88C} {763, 88}

\bibitem[\protect\citeauthoryear{{Childress} et~al.,}{{Childress}
  et~al.}{2013}]{childress13}
{Childress} M.,  et~al., 2013, \mndoi [\apj] {10.1088/0004-637X/770/2/108},
  \href {http://adsabs.harvard.edu/abs/2013ApJ...770..108C} {770, 108}

\bibitem[\protect\citeauthoryear{{Conley} et~al.,}{{Conley}
  et~al.}{2011}]{conley11}
{Conley} A.,  et~al., 2011, \mndoi [\apjs] {10.1088/0067-0049/192/1/1}, \href
  {http://adsabs.harvard.edu/abs/2011ApJS..192....1C} {192, 1}

\bibitem[\protect\citeauthoryear{{Contreras} et~al.,}{{Contreras}
  et~al.}{2010}]{contreras10}
{Contreras} C.,  et~al., 2010, \mndoi [\aj] {10.1088/0004-6256/139/2/519},
  \href {http://adsabs.harvard.edu/abs/2010AJ....139..519C} {139, 519}

\bibitem[\protect\citeauthoryear{{Ellis} et~al.,}{{Ellis}
  et~al.}{2008}]{ellis08}
{Ellis} R.~S.,  et~al., 2008, \mndoi [\apj] {10.1086/524981}, \href
  {http://adsabs.harvard.edu/abs/2008ApJ...674...51E} {674, 51}

\bibitem[\protect\citeauthoryear{{Foreman-Mackey}, {Hogg}, {Lang}  \&
  {Goodman}}{{Foreman-Mackey} et~al.}{2013}]{foreman13}
{Foreman-Mackey} D.,  {Hogg} D.~W.,  {Lang} D.,   {Goodman} J.,  2013, \mndoi
  [\pasp] {10.1086/670067}, \href
  {http://adsabs.harvard.edu/abs/2013PASP..125..306F} {125, 306}

\bibitem[\protect\citeauthoryear{{Frieman}, {Turner}  \& {Huterer}}{{Frieman}
  et~al.}{2008}]{frieman08}
{Frieman} J.~A.,  {Turner} M.~S.,   {Huterer} D.,  2008, \mndoi [\araa]
  {10.1146/annurev.astro.46.060407.145243}, \href
  {http://adsabs.harvard.edu/abs/2008ARA%26A..46..385F} {46, 385}

\bibitem[\protect\citeauthoryear{{Goobar} \& {Leibundgut}}{{Goobar} \&
  {Leibundgut}}{2011}]{goobar11}
{Goobar} A.,  {Leibundgut} B.,  2011, \mndoi [Annual Review of Nuclear and
  Particle Science] {10.1146/annurev-nucl-102010-130434}, \href
  {http://adsabs.harvard.edu/abs/2011ARNPS..61..251G} {61, 251}

\bibitem[\protect\citeauthoryear{{Gupta} et~al.,}{{Gupta}
  et~al.}{2016}]{gupta16}
{Gupta} R.~R.,  et~al., 2016, preprint, \href
  {http://adsabs.harvard.edu/abs/2016arXiv160406138G} {} (\mn@eprint {arXiv}
  {1604.06138})

\bibitem[\protect\citeauthoryear{{Guy} et~al.,}{{Guy} et~al.}{2010}]{guy10}
{Guy} J.,  et~al., 2010, \mndoi [\aap] {10.1051/0004-6361/201014468}, \href
  {http://adsabs.harvard.edu/abs/2010A%26A...523A...7G} {523, A7}

\bibitem[\protect\citeauthoryear{{Hamuy} et~al.,}{{Hamuy}
  et~al.}{2006}]{hamuy06}
{Hamuy} M.,  et~al., 2006, \mndoi [\pasp] {10.1086/500228}, \href
  {http://adsabs.harvard.edu/abs/2006PASP..118....2H} {118, 2}

\bibitem[\protect\citeauthoryear{{Hicken} et~al.,}{{Hicken}
  et~al.}{2009}]{hicken09}
{Hicken} M.,  et~al., 2009, \mndoi [\apj] {10.1088/0004-637X/700/1/331}, \href
  {http://adsabs.harvard.edu/abs/2009ApJ...700..331H} {700, 331}

\bibitem[\protect\citeauthoryear{{Hicken} et~al.,}{{Hicken}
  et~al.}{2012}]{hicken12}
{Hicken} M.,  et~al., 2012, \mndoi [\apjs] {10.1088/0067-0049/200/2/12}, \href
  {http://adsabs.harvard.edu/abs/2012ApJS..200...12H} {200, 12}

\bibitem[\protect\citeauthoryear{{Holtzman} et~al.,}{{Holtzman}
  et~al.}{2008}]{holtzman08}
{Holtzman} J.~A.,  et~al., 2008, \mndoi [\aj] {10.1088/0004-6256/136/6/2306},
  \href {http://adsabs.harvard.edu/abs/2008AJ....136.2306H} {136, 2306}

\bibitem[\protect\citeauthoryear{{Howell} et~al.,}{{Howell}
  et~al.}{2005}]{howell05}
{Howell} D.~A.,  et~al., 2005, \mndoi [\apj] {10.1086/497119}, \href
  {http://adsabs.harvard.edu/abs/2005ApJ...634.1190H} {634, 1190}

\bibitem[\protect\citeauthoryear{{Jha}, {Riess}  \& {Kirshner}}{{Jha}
  et~al.}{2007}]{jha07}
{Jha} S.,  {Riess} A.~G.,   {Kirshner} R.~P.,  2007, \mndoi [\apj]
  {10.1086/512054}, \href {http://adsabs.harvard.edu/abs/2007ApJ...659..122J}
  {659, 122}

\bibitem[\protect\citeauthoryear{{J{\"o}nsson} et~al.,}{{J{\"o}nsson}
  et~al.}{2010}]{jonsson10}
{J{\"o}nsson} J.,  et~al., 2010, \mndoi [\mnras]
  {10.1111/j.1365-2966.2010.16467.x}, \href
  {http://adsabs.harvard.edu/abs/2010MNRAS.405..535J} {405, 535}

\bibitem[\protect\citeauthoryear{{Joyce}, {Lombriser}  \& {Schmidt}}{{Joyce}
  et~al.}{2016}]{joyce16}
{Joyce} A.,  {Lombriser} L.,   {Schmidt} F.,  2016, preprint, \href
  {http://adsabs.harvard.edu/abs/2016arXiv160106133J} {} (\mn@eprint {arXiv}
  {1601.06133})

\bibitem[\protect\citeauthoryear{{Kelly}, {Filippenko}, {Burke}, {Hicken},
  {Ganeshalingam}  \& {Zheng}}{{Kelly} et~al.}{2015}]{kelly15}
{Kelly} P.~L.,  {Filippenko} A.~V.,  {Burke} D.~L.,  {Hicken} M.,
  {Ganeshalingam} M.,   {Zheng} W.,  2015, \mndoi [Science]
  {10.1126/science.1261475}, \href
  {http://adsabs.harvard.edu/abs/2015Sci...347.1459K} {347, 1459}

\bibitem[\protect\citeauthoryear{{Lidman} et~al.,}{{Lidman}
  et~al.}{2013}]{lidman13}
{Lidman} C.,  et~al., 2013, \mndoi [\pasa] {10.1017/pasa.2012.001}, \href
  {http://adsabs.harvard.edu/abs/2013PASA...30....1L} {30, 1}

\bibitem[\protect\citeauthoryear{{Mannucci}, {Della Valle}  \&
  {Panagia}}{{Mannucci} et~al.}{2006}]{mannucci06}
{Mannucci} F.,  {Della Valle} M.,   {Panagia} N.,  2006, \mndoi [\mnras]
  {10.1111/j.1365-2966.2006.10501.x}, \href
  {http://adsabs.harvard.edu/abs/2006MNRAS.370..773M} {370, 773}

\bibitem[\protect\citeauthoryear{{Mosher} et~al.,}{{Mosher}
  et~al.}{2014}]{mosher14}
{Mosher} J.,  et~al., 2014, \mndoi [\apj] {10.1088/0004-637X/793/1/16}, \href
  {http://adsabs.harvard.edu/abs/2014ApJ...793...16M} {793, 16}

\bibitem[\protect\citeauthoryear{{Perlmutter} et~al.,}{{Perlmutter}
  et~al.}{1999}]{perlmutter99}
{Perlmutter} S.,  et~al., 1999, \mndoi [\apj] {10.1086/307221}, \href
  {http://adsabs.harvard.edu/abs/1999ApJ...517..565P} {517, 565}

\bibitem[\protect\citeauthoryear{{Phillips}}{{Phillips}}{1993}]{phillips93}
{Phillips} M.~M.,  1993, \mndoi [\apjl] {10.1086/186970}, \href
  {http://adsabs.harvard.edu/abs/1993ApJ...413L.105P} {413, L105}

\bibitem[\protect\citeauthoryear{{Riess} et~al.,}{{Riess}
  et~al.}{1998}]{riess98}
{Riess} A.~G.,  et~al., 1998, \mndoi [\aj] {10.1086/300499}, \href
  {http://adsabs.harvard.edu/abs/1998AJ....116.1009R} {116, 1009}

\bibitem[\protect\citeauthoryear{{Riess} et~al.,}{{Riess}
  et~al.}{1999}]{riess99}
{Riess} A.~G.,  et~al., 1999, \mndoi [\aj] {10.1086/300738}, \href
  {http://adsabs.harvard.edu/abs/1999AJ....117..707R} {117, 707}

\bibitem[\protect\citeauthoryear{{Riess} et~al.,}{{Riess}
  et~al.}{2007}]{riess07}
{Riess} A.~G.,  et~al., 2007, \mndoi [\apj] {10.1086/510378}, \href
  {http://adsabs.harvard.edu/abs/2007ApJ...659...98R} {659, 98}

\bibitem[\protect\citeauthoryear{{Sako} et~al.,}{{Sako} et~al.}{2011}]{sako11}
{Sako} M.,  et~al., 2011, \mndoi [\apj] {10.1088/0004-637X/738/2/162}, \href
  {http://adsabs.harvard.edu/abs/2011ApJ...738..162S} {738, 162}

\bibitem[\protect\citeauthoryear{{Sako} et~al.,}{{Sako} et~al.}{2014}]{sako14}
{Sako} M.,  et~al., 2014, preprint, \href
  {http://adsabs.harvard.edu/abs/2014arXiv1401.3317S} {} (\mn@eprint {arXiv}
  {1401.3317})

\bibitem[\protect\citeauthoryear{{Stritzinger} et~al.,}{{Stritzinger}
  et~al.}{2011}]{stritzinger11}
{Stritzinger} M.~D.,  et~al., 2011, \mndoi [\aj] {10.1088/0004-6256/142/5/156},
  \href {http://adsabs.harvard.edu/abs/2011AJ....142..156S} {142, 156}

\bibitem[\protect\citeauthoryear{{Sullivan} et~al.,}{{Sullivan}
  et~al.}{2010}]{sullivan10}
{Sullivan} M.,  et~al., 2010, \mndoi [\mnras]
  {10.1111/j.1365-2966.2010.16731.x}, \href
  {http://adsabs.harvard.edu/abs/2010MNRAS.406..782S} {406, 782}

\bibitem[\protect\citeauthoryear{{Tripp}}{{Tripp}}{1998}]{tripp98}
{Tripp} R.,  1998, \aap, \href
  {http://adsabs.harvard.edu/abs/1998A%26A...331..815T} {331, 815}

\bibitem[\protect\citeauthoryear{Uddin}{Uddin}{2016}]{uddinphd}
Uddin S.~A.,  2016, PhD thesis, Swinburne University of Technology

\bibitem[\protect\citeauthoryear{{Walker} et~al.,}{{Walker}
  et~al.}{2011}]{walker11}
{Walker} E.~S.,  et~al., 2011, \mndoi [\mnras]
  {10.1111/j.1365-2966.2010.17519.x}, \href
  {http://adsabs.harvard.edu/abs/2011MNRAS.410.1262W} {410, 1262}

\bibitem[\protect\citeauthoryear{{Yuan} et~al.,}{{Yuan} et~al.}{2015}]{yuan15}
{Yuan} F.,  et~al., 2015, \mndoi [\mnras] {10.1093/mnras/stv1507}, \href
  {http://adsabs.harvard.edu/abs/2015MNRAS.452.3047Y} {452, 3047}

\bibitem[\protect\citeauthoryear{{Zheng} et~al.,}{{Zheng}
  et~al.}{2008}]{zheng08}
{Zheng} C.,  et~al., 2008, \mndoi [\aj] {10.1088/0004-6256/135/5/1766}, \href
  {http://adsabs.harvard.edu/abs/2008AJ....135.1766Z} {135, 1766}

\makeatother
\end{thebibliography}

\end{document}